\documentclass[aps,prd,superscriptaddress,amsfonts,amssymb,amsmath,eqsecnum,nofootinbib,twocolumn,floatfix]{revtex4}
 \usepackage{color,graphicx}
 \usepackage[utf8]{inputenc}
  \usepackage[T2A]{fontenc}
 \usepackage[english]{babel}
 \definecolor{darkblue}{rgb}{0,0,0.7}
\definecolor{darkred}{rgb}{0.7,0,0}
\definecolor{darkgreen}{rgb}{0,0.4,0}
\usepackage[unicode, colorlinks, citecolor=darkblue, linkcolor=darkred, urlcolor=blue]{hyperref}

  \renewcommand{\Re}{\mathfrak{Re}}
\renewcommand{\Im}{\mathfrak{Im}}
\allowdisplaybreaks
\begin{document}

\author{Alexandr Karpenko}
\affiliation{Faculty of Physics, M.V. Lomonosov Moscow State University, Leninskie Gory, Moscow 119991,  Russia}

\author{Mikhail Korobko}
\affiliation{Institut fur Laserphysik \& Zentrum fur Optische Quantentechnologien, Universitat Hamburg, Luruper Chaussee 149,22761 Hamburg, Germany}

\author{Sergey P. Vyatchanin}
\affiliation{Faculty of Physics, M.V. Lomonosov Moscow State University, Leninskie Gory, Moscow 119991,  Russia}
\affiliation{Quantum Technology Centre, M.V. Lomonosov Moscow State University, Leninskie Gory, Moscow 119991,  Russia}

\date{\today}
	
\title{Enhanced optomechanical interaction in the unbalanced interferometer}

\begin{abstract}
Quantum optomechanical systems enable the study of fundamental questions on quantum nature of massive objects. For that a strong coupling between light and mechanical motion is required, which presents a challenge for massive objects. In particular large interferometric sensors with low frequency oscillators are difficult to bring into quantum regime. Here we propose a modification of the Michelson-Sagnac interferometer, which allows to boost the optomechanical coupling strength. This is done by unbalancing the central beam-splitter of the interferometer, allowing to balance two types of optomechanical coupling present in the system: dissipative and dispersive. We analyse two different configurations, when the optomechanical cavity is formed by the mirror for the laser pump field (power-recycling), and by the mirror for the signal field (signal-recycling). We show that the imbalance of the beam splitter allows to dramatically increase the optical cooling of the test mass motion. We also formulate the conditions for observing quantum radiation-pressure noise and ponderomotive squeezing. Our configuration can serve as the basis for more complex modifications of the interferometer that would utilize the enhanced coupling strength. This will allow to efficiently reach quantum state of large test masses, opening the way to studying fundamental aspects of quantum mechanics and experimental search for quantum gravity.
\end{abstract}

\maketitle

\section{Introduction}
Modern quantum technology allows measurements of small forces and displacements with unprecedented precision.
This is achieved by utilizing the power of interaction between the test mass and the electromagnitic field that is used to probe it.
This interaction is amplified by embedding the test mass into a cavity.
Quantum optomechanical devices are used for a wide range of applications: from micro-sensors\cite{Young2018, Taylor2016} and applied metrological device\,\cite{Nolte2012, Krause2012}, to fundamental probes for macroscopic quantum effects\,\cite{Yu2020, gonzalez2021levitodynamics,  schnabel2022macroscopic} or cosmology\,\cite{backes2021quantum} to large-scale gravitational-wave detectors\cite{aLIGO2013,aLIGO2015,MartynovPRD16,AserneseCQG15, DooleyCQG16,AsoPRD13,Acernese2019,Tse2019}.

The interaction between the light and the mechanical oscillator leads to quantum radiation pressure noise (QRPN), exerted on the mechanical oscillator\,\cite{Caves1981}.
Optical cavity provides the coherent feedback mechanism, which allows to manipulate the state of the mechanical object and its interaction with the state of the light field\,\cite{aspelmeyerQuantumOptomechanics2012}.
This makes cavity optomechanical systems particularly versatile as sensors.

From the fundamental perspective, the first step in creating a good sensor is bringing it into quantum regime.
There are several hallmarks of achieving that: the ability to cool the mechanical oscillator to its motional ground state, observation of QRPN, and observation of quantum correlations between the oscillator and the light field.
Some configurations, where microscopic objects such as photonic crystals or nano-membranes are measured, have been spectacularly successful in reaching quantum regime\,\cite{mason2019continuous}.
Others, which involve low-frequency and high-mass oscillators, have much more stringent requirements on the particular setup, and only a few have actually demonstrated quantum features\,\cite{cripe2020quantum,yap2020broadband}, including LIGO and Virgo detectors with 40\,kg test masses\,\cite{Yu2020,acernese2020quantum,whittle2021approaching}.


%

In this paper we study the Michelson-Sagnac cavity-enhanced interferometer.
This interferometric design has unique features that make it stand out as an effective tool for low-frequency and high-mass optomechanical sensors\,\cite{SawadskyPRL2015}.
Generally, there are three ways how light and mechanical oscillators couple to each other: dispersively, coherently and dissipatively.
In dispersive coupling, mirror displacement causes the change in the frequency of the cavity normal mode\,\cite{aspelmeyerQuantumOptomechanics2012}.
In coherent coupling, mirror displacement causes the coherent exchange between the two cavity modes\,\cite{li2019coherent}.
In dissipative coupling, mirror displacement changes the cavity relaxation rate. Dissipative coupling was  proposed theoretically \cite{ElstePRL2009} and confirmed experimentally \cite{LiPRL2009,WeissNJP2013,WuPRX2014, HryciwOpt2015} about a decade ago. This phenomenon was investigated in numerous optomechanical systems, including the FP interferometer \cite{LiPRL2009,WeissNJP2013,WuPRX2014, HryciwOpt2015}, the Michelson-Sagnac interferometer (MSI) \cite{XuerebPRL2011, TarabrinPRA2013, SawadskyPRL2015, 16a1PRAVyMa, 19a1JPhyBNaVy, 20PRAKaVy}, and ring resonators \cite{HuangPRA2010,HuangPRA2010b}. It was demonstrated that an optomechanical transducer based on dissipative coupling  allows realizing quantum speed meter which, in turn, helps to avoid Standard Quantum Limit (SQL) \cite{16a1PRAVyMa}. Recently it was shown that {\em combination} of dispersive and dissipative coupling can be useful to surpass SQL and to create optical rigidity even at resonance pump \cite{20PRAKaVy, 22PRAKaVy}.

In this paper we propose the new configuration of MSI, which allows to enhance its sensitivity.
This configuration takes advantage of unbalanced beam splitter in the interferometer.
Usually, unbalanced beam splitter is seen as hindrance in interferometric experiments. 
In our setup, this imbalance allows to optimize the combination of dissipative and dispersive coupling to enhance their strength.
We investigate this combination of dissipative and dispersive coupling and show how it can be used for parametric cooling of a mechanical oscillator even on cavity resonance, as well as observation of QRPN and ponderomotive squeezing. 
We study two different configurations, where the cavity is formed for the pump be\textbf{}am (power-recycling) and for the signal (signal-recycling), compare their performance and find the optimal operating points.
We demonstrate that large imbalance in the beam-splitter allows to dramatically increase optical rigidity in the system, which, in turn, allows to increase parametric cooling of the test mass. 
We also formulate the conditions for observing QRPN, ponderomotive squeezing and efficient cooling in a table-top experiment with micro-mechanical membrane.

We study different parameter regimes of operating the interferometer with imbalanced beam-splitter, and lay the groundwork for improved design of quantum optomechanical experiments with heavy test masses, which will feature both signal and power recycling, as well as detuned operation.
Large mass is crucial for many tests of gravity\,\cite{westphal2021measurement}, fundamental aspects of quantum mechanics\,\cite{marshall2003towards} and the experimental search for quantum gravity\,\cite{datta2021signatures}.

\begin{figure}
	\includegraphics[width=0.5\textwidth]{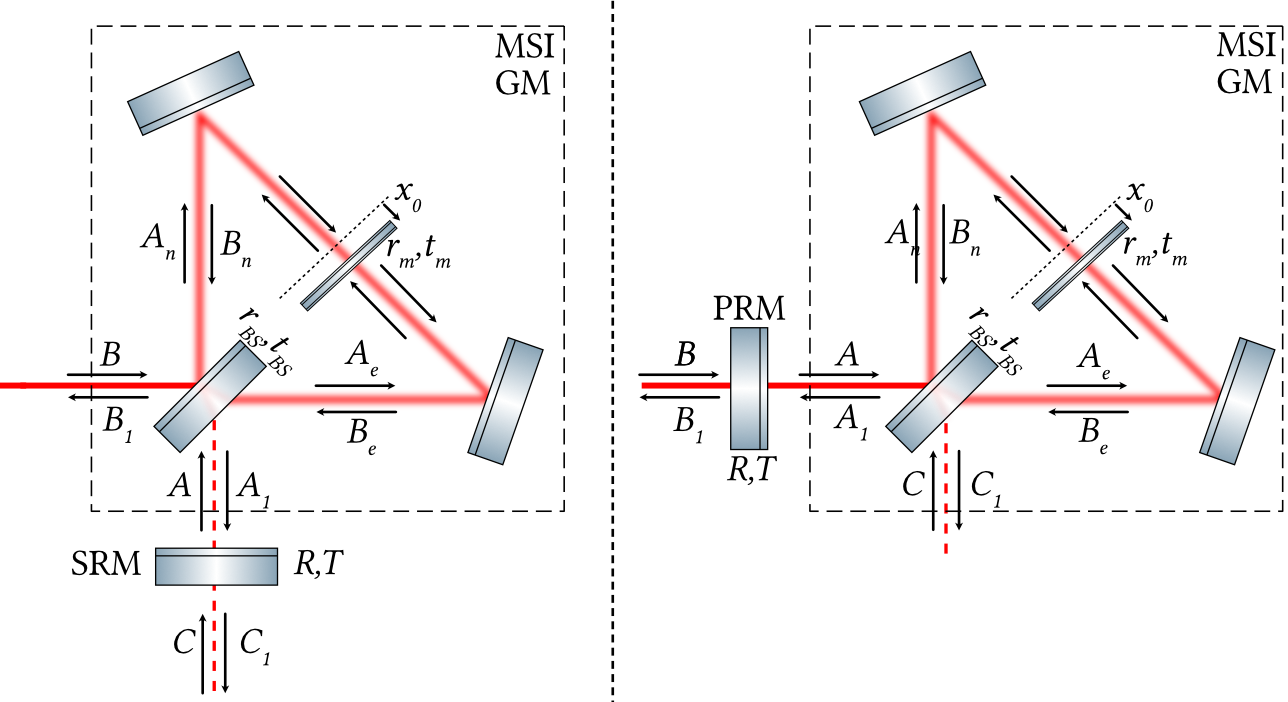}
	\caption{SRM cavity (left) and PRM cavity (right). Michelson-Sagnac interferometer with movable mirror $M$ (which is a mass of probe oscillator) is a generalized {\em input} mirror (GM)  of FP cavity.
 Central beam splitter can be imbalanced and has a the amplitude reflection and transmission coefficients $r_{BS}, t_{BS}$.
 Partially transmissive movable mirror (M) moves under the effect of thermal and radiation-pressure noise. 
 Its average displacement $x_0$ from the center of the interferometer influenced the balance between the dissipative and dispersive optomehcanical coupling.
 Output wave $C_1$ contains information about the motion of the test mirror and can be read out with balanced homodyne detection.}\label{CavSRM}
\end{figure}


\section{Michelson-Sagnac interferometer}
We study the Michelson-Sagnac interferometer (MSI), which is enhanced with a cavity on either the signal, or the pump port.
This way, the MSI acts as a generalised mirror, forming a cavity with a signal recylcing mirror (SRM) or power recycling mirror (PRM), see Fig.~\ref{CavSRM}.
In both schemes an optical mode with eigenfrequency $\omega_0$ is pumped with resonant light. The optical mode is coupled to the mechanical oscillator with mass $m$, eigen frequency $\omega_m$ and position $x$.  
The output $C_1$ carries information about the displacement $x$, and is measured on the homodyne detector.
The incoming vacuum field $C$ could be squeezed to enhance the sensitivity of the interferometer\,\cite{schnabel2017squeezed}.

We follow\,\cite{SawadskyPRL2015, SawadskyArxiv2015} in describing the MSI. 
The transmission of the MSI depends on the reflectivity of the movable mirror $r_m$, its displacement $x_0$ relative to the equal distance from the beam splitter, and the imbalance of the beam splitter $\epsilon$. 
The MSI can be viewed as a generalized mirror with transmission $T_{\rm MSI}$ and reflection $R_{\rm MSI}$\cite{SawadskyArxiv2015}:
\begin{subequations}
\label{TRmsi}
\begin{align}
 T_{\rm MSI}& = r_m\sqrt{1-\epsilon^2}\sin 2kx_0 - t_m\epsilon,\\
R_{\rm MSI} &= r_m\left(\cos 2kx_0 +i\epsilon \sin 2kx_0\right) +it_m\sqrt{1-\epsilon^2},
\end{align}
\end{subequations}
where $k=\omega_{0}/c$ is the wave number, $t_m,\, r_m$ are amplitude transmittance and reflectivity of mirror $M$, the imbalance $\epsilon$ is defined with respect to the balanced case, and connects to the reflectivity and transmissivity of the beam splitter:
\begin{align}
 \label{eps}
 r_{BS}^2 &= \frac{1+\epsilon}{ 2}, \quad 
    t_{BS}^2= \frac{1-\epsilon}{2}.
\end{align}
Displacement $x$ changes optical eigenfrequency $\omega_0$ and optical relaxation rate $\gamma_0$:
\begin{subequations}
 \label{xieta}
 \begin{align}
  \omega&=\omega_{0}(1+\xi x),\quad 
  \xi=  \frac{t_m T_{\rm MSI}+\epsilon}{L|R_{\rm MSI}|^2},\\
  \gamma &=\gamma_{0} (1+\eta x),\quad \eta = \frac{4kr_m\sqrt{1-\epsilon^2}\cos 2kx_0}{T_{\rm MSI}},
 \end{align}
\end{subequations}
where $L$ is the effective cavity length, and $\xi,\, \eta$ are coefficients of dispersive and dissipative coupling.
These coefficients can be derived from \eqref{TRmsi} using definitions:
\begin{align}
\label{etaxi}
 \eta &\equiv 2\, \frac{\partial_x T_{\rm MSI}}{T_{\rm MSI}},
   \quad
 \xi \equiv\frac{1}{\omega_0 \tau} \cdot
  \Im\left(\frac{\partial_x R_{\rm MSI}}{R_{\rm MSI}}\right).
\end{align}
The cavity has two relaxation rates
\begin{align}
 \gamma_0 &= \frac{T_{\rm MSI}^2}{\tau},\quad \gamma_1 = \frac{T^2}{\tau},
\end{align}
where $\gamma_{0}$ describes relaxation due to MSI and $\gamma_1$ --- due to SRM or PRM on Fig.~\ref{CavSRM}, and $\tau=2L/c$. The total relaxation rate of field inside cavity is equal to $(\gamma_1+\gamma_0)$.

From \eqref{xieta} we see that in case of symmetric beam splitter ($\epsilon=0$) and completely reflecting mirror $M$ ($t_m=0$) dispersive coupling is absent $\xi=0$ and only dissipative coupling exists ($\eta\ne 0$). However, for the non-symmetric beam splitter dispersive coupling can be rather large even in case of $t_m=0$: $\xi\sim \epsilon/L$.  This is one of the main findings of the paper. Below we analyze some examples of application of large large dispersive coupling in combination with dissipative one.

\section{Parametric cooling}
\label{BA} 

Thermal dissipation in mechanical oscillator leads to thermal noise, exciting its motion, as described by the fluctuation-dissipation theorem.
The amount of thermal phonons in steady state is defined by the temperature of the environment:
\begin{equation}
    n_{T} = \frac{k_B T_0}{\hbar \omega_m}.
\end{equation}
These thermal phonon prevent from observing quantum effects on the mechanical oscillator. 
The mean thermal occupation should be brought below zero in order to perform quantum experiments. 
It could be done by reducing the temperature $T_0$, performing conditional measurements\,\cite{Miao2010}, or cooling the oscillator by actively extracting thermal energy.
This can be done by introducing optical feedback in the system and depositing thermal energy into optical field.
Such optical feedback in the cavity can lead to the radiation pressure modifying the mechanical response of the oscillator, introducing so-called optical rigidity\,\cite{braginski1967ponderomotive,khalili2016quantum}.
This rigidity causes the shift in the mechanical frequency and the mechanical relaxation rate.

Usually, optical rigidity in optical cavities appears only when the laser is detuned from cavity resonance.
However, recently it was established that the combination of dissipative and dispersive coupling causes optical rigidity even at resonance pump \cite{SawadskyPRL2015, 20PRAKaVy, 22PRAKaVy}. 
Here we compute the optical spring frequency for SRM and PRM cavities:
\begin{align}
  \label{OmSRM}
 \text{SRM: } \omega_{os}^2 &=
        - \frac{W_{in} \gamma_0^2 \eta \xi}{m\gamma_+^2(\gamma_+ - i\Omega)},\\
 \label{OmPRM}
 \text{PRM: }\omega_{os}^2 &=
        - \frac{W_{in}\gamma_1\gamma_0 \eta \xi }{m\gamma_+^2(\gamma_+ - i\Omega)},\\
 \label{gammaPM}
    \gamma_+ &= \frac{\gamma_1+\gamma_0}{2},\quad 
        \gamma_-=\frac{\gamma_1-\gamma_0}{2}.
\end{align}
where $W_{in}$ is input light power, $m$ is the mass of a mechanical oscillator.
Dynamical characteristics of the oscillator (eigenfrequency $\omega_m$ and relaxation rate $\kappa_m$) are transformed under the action of optical rigidity:
\begin{align}
\label{OmM}
 \omega_M^2 &\equiv \omega_m^2 +\Re(\omega_{os}^2),\quad
   \kappa_M = \kappa_m  - \frac{\Im(\omega_{os}^2)}{\Omega},
\end{align}
See details in Appendices \ref{aSRM} and \ref{aPRM}.

If introduced rigidity is negative, $\Re(\omega_{os}^2)<0$, then introduced relaxation is positive, $\Im(\omega_{os}^2) > 0$, and it can be used for parametric cooling. Indeed, introduction of low noise optic relaxation decreases mean thermal energy of the mechanical oscillator  from initial $k_B T_0$ to $k_B T_\text{eff}$ ($k_B$ is Boltzmann constant):
\begin{subequations}
\begin{align}
  \label{Teff}
 n_T \simeq&  \frac{\kappa_m}{\kappa_M} \frac{k_B T_0}{\hslash \omega_M}
    + \left(\frac{S_{LP}}{2\kappa_M}\right),\\ T_\text{eff}=&\frac{n_T\hslash \omega_M}{k_B},\\
  \label{SSRM}
   S_{LP}^{SRM}&\simeq  \frac{\gamma_0^2}{2\omega_M}\frac{\gamma_+^2}{ (\gamma_+^2 +\omega_M^2)}\\
   &\quad \times \left( \mathcal{H}^2\left[1+\frac{\gamma_1}{\gamma_0}\right]
   + \mathcal{X}^2\left[\frac{\gamma_1}{\gamma_0}\, 
	 + \frac{\omega_M^2}{\gamma_+^2}\right]\right),\\
   \label{SPRM}
   S_{LP}^{PRM} & \simeq \frac{\gamma_1^2}{2\omega_M}\left(
  \mathcal{H}^2\frac{\gamma_+^2}{\gamma_+^2 +\omega_M^2}\left[1+\frac{\gamma_0}{\gamma_1}\right]
   + \mathcal{X}^2\left[\frac{\gamma_0}{\gamma_1}\right]\right),\\
  \label{PQ}
   \mathcal{X}&= \eta \sqrt\frac{W_{in}}{2m\omega_0 \gamma_+^2},\quad 
     \mathcal{H} = \xi \sqrt \frac{2 \omega_0 W_{in}}{m \gamma_+^4} ,
\end{align}
\end{subequations}
where $S_{LP}$ is a spectral density of QRPN, and $\mathcal{X},\mathcal{H}$ are the normalized optomechanical coupling rates, see details in Appendices \ref{aSRM} and \ref{aPRM}. Here we assume conditions 
\begin{align}
\label{cond2}
   k_BT_0\gg \hslash \omega_M,\quad \kappa_m,\kappa_M\ll \omega_m,\quad \omega_M \ll \gamma_+.  
   \end{align}

Since optical rigidity (and optical relaxation rate) is proportional to $\xi\eta$, we have to maximize this product in order to increase cooling. From Eq.\eqref{xieta} we derive the optimal value for the imbalance of the beam splitter $\epsilon_{opt}$, for which $\kappa_M$ takes the maximum value:
\begin{align}
\label{coolingmax}
 \epsilon_{opt} = \frac{r_m}{\sqrt{2}}\sqrt{1-\tau\gamma_0}-t_m\sqrt{\tau\gamma_0}.
\end{align}
At the same time, $r_m$ must be taken large enough to have sufficient resonance gain for optical power (i.e. relaxation rate $\gamma_0$ has to be small).

\begin{figure}
\includegraphics[width=0.5\textwidth]{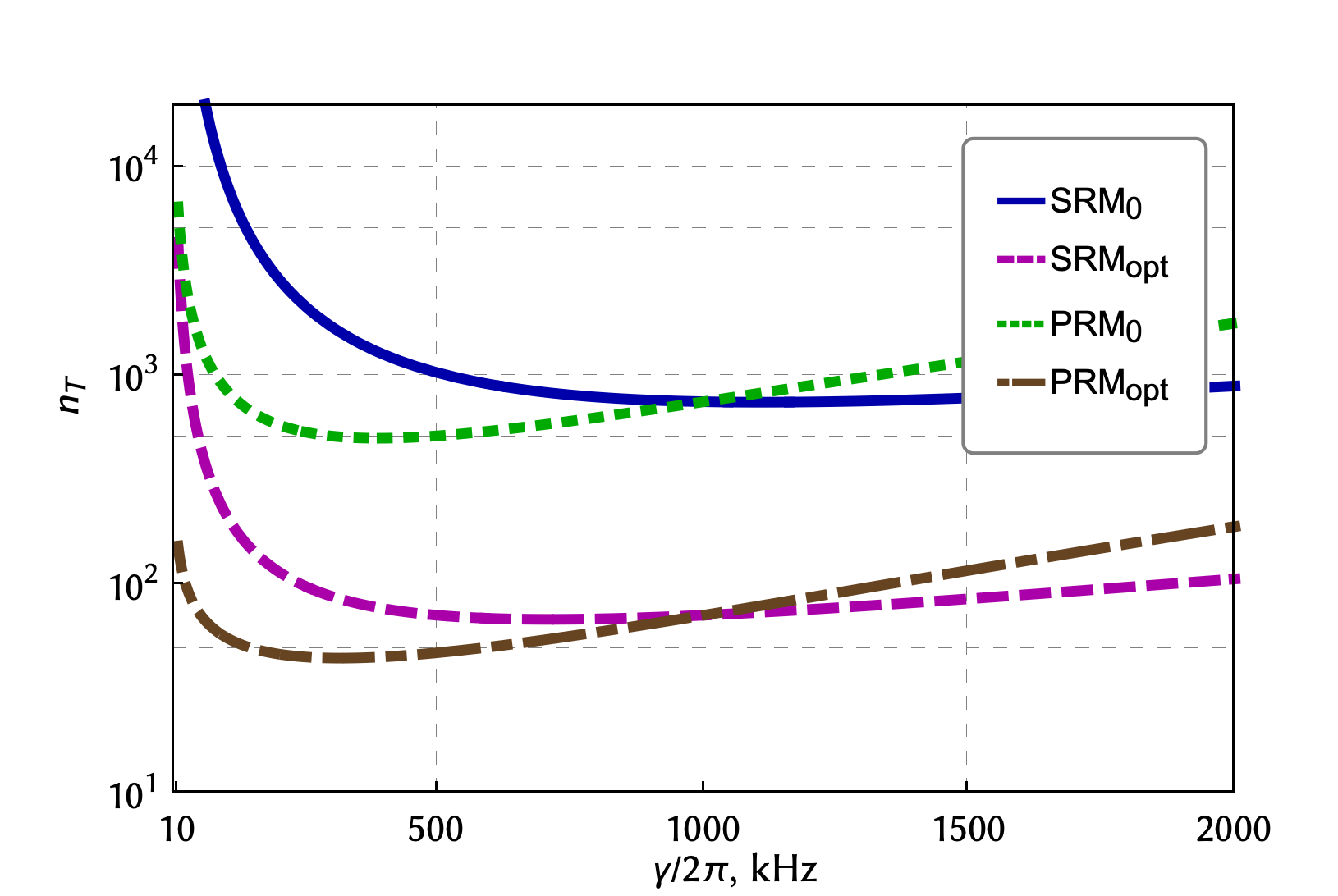}
	\caption{Cooling of a micro-mechanical membrane in the MSI. The plot shows the dependence of the thermal photons number $n_T$ \eqref{Teff} of a mechanical oscillator on the MSI bandwidth $\gamma_0$ for SRM and PRM schemes at $\epsilon=0$ (symmetric beam splitter) and $\epsilon_{opt}$ (optimally asymmetric beam splitter).  To construct the dependencies, we used the following values of the system parameters: the bandwidth of the SRM (PRM) $\gamma_1/2\pi = 10^6 Hz$, the reflection coefficient of the membrane $r_m^2 = 0.5$ for $\text{SRM}_0, \text{PRM}_0$ and $r_m^2 = 0.98$ for $\text{SRM}_{\text{opt}}, \text{PRM}_{\text{opt}}$. The values of other parameters are taken from the Table \ref{Table}.}
 \label{cooling}
\end{figure}

\begin{figure}
\includegraphics[width=0.5\textwidth]{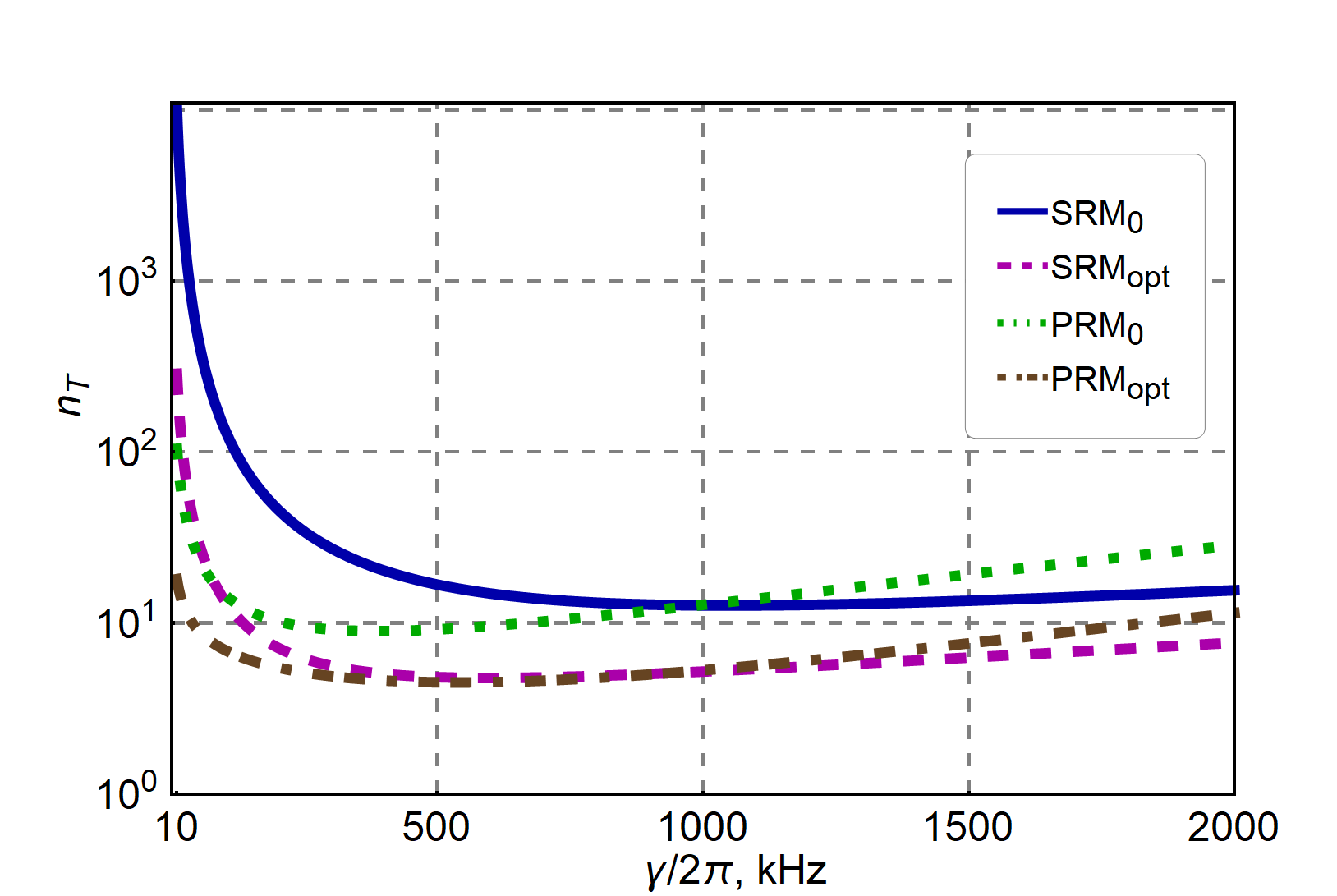}
	\caption{Cooling of a micro-mechanical membrane in the MSI. The plot shows the dependence of the thermal photons number $n_T$ \eqref{Teff} of a mechanical oscillator on the MSI bandwidth $\gamma_0$ for SRM and PRM schemes at $\epsilon=0$ (symmetric beam splitter) and $\epsilon=0.15$.  To construct the dependencies, we used the following values of the system parameters: Input power $P_{in} = 1 W$, Quality factor is $10^7$, the bandwidth of the SRM (PRM) $\gamma_1/2\pi = 10^6 Hz$, the reflection coefficient of the membrane $r_m^2 = 0.5$ for $\text{SRM}_0, \text{PRM}_0$ and $r_m^2 = 0.98$ for $\text{SRM}_{\text{opt}}, \text{PRM}_{\text{opt}}$, 6dB squeezing of the field $C$ quadrature (for a scheme with SRM it is \eqref{SRMct2b}, and for a scheme with PRM it is \eqref{PRMct2b}). The values of other parameters are taken from the Table \ref{Table}.}
 \label{cooling1}
\end{figure}

Fig.~\ref{cooling} shows the dependence of the thermal phonon number $n_{T}$ of a mechanical oscillator on the MSI bandwidth $\gamma_0$ for SRM and PRM schemes at $\epsilon=0$ and $\epsilon_{opt}$. We see that the use of an asymmetric beam splitter ($\epsilon \neq 0$) can significantly increase the cooling of the mechanical oscillator in both schemes. This is due to a sharp increase in the coefficient of dispersive coupling $\xi$ \eqref{xieta} when we use  non-symmetric beam splitter. 

We also see that a scheme with PRM gives a slightly better result than a scheme with SRM at $\gamma_0 < \gamma_1$, since the maximal light power inside the resonator is reached at $\gamma_0 \ll \gamma_1$ for PRM, and at $\gamma_0 = \gamma_1$ for SRM. This maximum power in the scheme with PRM is approximately 4 times greater than in the scheme with SRM. This follows from Eq.\eqref{AB1} and Eq.\eqref{AB2} for the amplitude of the field inside the resonator for the case with SRM and PRM.

Using schemes with the parameters from the table \ref{Table} we can cool the mechanical oscillator to a few tens of thermal phonons. But how do we cool the mechanical oscillator down to its ground state? For this we can increase the laser power and the quality factor of the oscillator by a factor of 10 ($P_{in} = 1 W$, $Q=10^7$). With these parameters, the thermal noise will be significantly reduced, and optical fluctuations will be the main contributor to the total noise. At large the imbalance of the beam splitter $\epsilon$, the power of optical fluctuations increases greatly due to the large difference in the powers of light in the arms of the MSI ($\epsilon \simeq 0.7$). But this does not mean that the unbalanced interferometer cannot be used to reduce optical fluctuations. Fig.~\ref{cooling1} shows the dependence of the thermal phonon number $n_{T}$ of a mechanical oscillator on the MSI bandwidth $\gamma_0$ for SRM and PRM schemes at $\epsilon=0$ and $\epsilon=0.15$, $P_{in} = 1 W$, $Q=10^7$ and 6dB squeezing of the field $C$ quadrature \footnote{It is necessary to squeeze the quadrature of the field $C$, on which the back action depends. For a scheme with SRM it is \eqref{SRMct2b}, and for a scheme with PRM it is \eqref{PRMct2b}.  }.  We see that at relatively small $\epsilon$ the cooling is better than at $\epsilon = 0$ and reaches the value of several phonon. This means that the unbalanced schemes allow cooling oscillators to several phonons and combat not only thermal but also optical noise.

\begin{table}[h]
 \caption{Parameters of a table-top MSI with a SiN membrane as mechanical oscillator.}\label{table1}
 \begin{tabular}{||c | c | c||}
 \hline
  \multicolumn{3}{||c||}{Membrane} \\
 \hline
 Mass, $m$ & 50 & $10^{-9}$ g\\
 Frequency, $\omega_m/2\pi$& 350 & $10^3$ Hz\\
 Quality factor $Q$ & $10^6$ & \\
 Temperature, $T$ & 20 & K$^\circ$ \\
 Thermal phonons number, $n_T$ & $1.2\cdot 10^6$ &\\
 Reflectivity $r^2_M$(power) & $8\dots 98$ & \% \\
 \hline
  \multicolumn{3}{||c||}{Cavity} \\
 \hline
 Arm length in MSI & 10 & cm \\
 Length of cavity & 5 & cm \\
 Bandwidth  (SRM or PRM), $\gamma_2/2\pi$ & $0.5 \dots 1$ & $10^6$ Hz\\
 Wave length, $\lambda= 2\pi c/\omega_0$ & 1550 & $10^{-9}$ m \\
 Input power $P_{in}$ & 100 & $10^{-3}$ W \\
 Squeezing & 6 & dB \\
 Anti-squeezing & 20 & dB \\
 \hline
 \end{tabular}
 \label{Table}
\end{table}


\section{Observation of fluctuation light pressure force}
\label{BackAction}
Vacuum fluctuations entering the interferometer lead to a fluctuating force on the test mass. 
Observing this quantum radiation pressure force is the necessary step in using the setup for quantum experiments.
In this chapter we will demonstrate how to use schemes on  Fig.~\ref{CavSRM} for observation of back action force, produced by fluctuating radiation pressure.

 Vacuum fluctuations enter the system as field $C$, and the laser light as field $B$ (see Fig.~\ref{CavSRM}).
 Laser field typically has significant contribution of classical frequency and amplitude noise.
 In a usual interferometer with balanced central beam splitter, this classical noise cancels due to destructive interference with itself.
 In our unbalanced setup, the contribution of classical noise would interfere with our observation of fluctuation radiation pressure force of the vacuum. This noise can be suppressed by the correct selection of $\gamma_0$ and $\gamma_1$. For a scheme with SRM, the radiation pressure force of the vacuum is proportional to $\sqrt{\gamma_1}$, and the radiation pressure force of the fluctuations of the laser light is proportional to $\sqrt{\gamma_0}$. It follows from this that we will be able to suppress the fluctuations of laser radiation if we make $\gamma_1\gg\gamma_0$. For a scheme with PRM the condition is correspondingly $\gamma_1\ll\gamma_0$.

 Another source of noise are thermal fluctuations. In  Eq.\,\eqref{Teff} the first term determines the total power of thermal fluctuations, and the second term determines the power of radiation pressure noise. Let's find their relation for the case with SRM:
 \begin{align}
 \label{TbaTt}
  \frac{\hbar\omega_m}{k_BT_0}\frac{S_{LP}}{2\kappa_m} \simeq \frac{\hbar\gamma_0\gamma_1\gamma_+^2}{4\kappa_mk_B T_0\left(\gamma_+^2+\omega_m^2\right)}\left(\mathcal{H}^2+\mathcal{X}^2\right)
 \end{align}
 Here we took into account that $\gamma_1\gg\gamma_0$. 

 Eq.\eqref{TbaTt} shows us how the powers of radiation pressure noise and thermal noise relate to each other. In order to reduce the influence of thermal fluctuations, we need to increase the ratio \eqref{TbaTt}. 
 We can do this by increasing factor $\mathcal{H}$, which in turn can be significantly increased for an asymmetric  beam splitter with non-zero coefficient $\epsilon$,
 since $\mathcal{H}\sim\xi\sim\epsilon$ \eqref{PQ}. Coefficient $\eta$ decreases with increase of $\epsilon$, and eventually becomes imaginary: 
  \begin{align}
\eta \sim \cos 2kx_0 = \sqrt{1-\left(\frac{\sqrt{\gamma_0\tau}+t_m\epsilon}{r_m\sqrt{1-\epsilon^2}}\right)^2}
 \end{align}

This defines the maximal value $\epsilon_{max}$, up to which we can increase the imbalance:
  \begin{align}
 \label{emax}
 \epsilon_{max} = r_m\sqrt{1-\gamma_0\tau}-t_m\sqrt{\gamma_0\tau}.
 \end{align}
From this equation one can see that $r_m$ needs to be chosen as high as possible.



At $\epsilon=\epsilon_{max}$ the coefficient of dissipative coupling $\eta=0$, that is, we get a system with a pure dispersive coupling. In optomechanical systems with pure dispersive coupling, information about the displacement of the mechanical generator is in the phase quadrature of the output field\,\cite{20PRAKaVy, 22PRAKaVy}. Therefore, for more effective observation of QRPN, the phase quadrature of the output field $C_1$ should be measured at $\epsilon=\epsilon_{max}$:
\begin{align}
c_{\phi} &= \frac{\sqrt{\gamma_1\gamma_0}}{\gamma_+-i\Omega}b_{\phi}+\frac{\gamma_-+i\Omega}{\gamma_+-i\Omega}c_{\phi} +\nonumber\\&+ \mathcal{H}\frac{\gamma_0\sqrt{\gamma_1}}{Z}\frac{\gamma_+^2}{\left(\gamma_+-i\Omega\right)^2}\left(\sqrt{\gamma_0}b_a+\sqrt{\gamma_1}c_a\right) -\nonumber\\&- \sqrt{2\mathcal{H}}\frac{\gamma_+}{\gamma_+-i\Omega}\frac{\sqrt{\gamma_0\gamma_1}}{Z}\sqrt{\omega_m}f_T,\\
Z &= \omega_M^2 - \Omega - i\Omega\kappa_M,\\
f_T &= \frac{F_T}{\sqrt{2\hbar m\omega_m}},\\ S_{F_T} &= 4m\kappa_m k_B T.
 \end{align}
 Here $f_T$ is a random thermal force normalized by SQL, definition of $c_\phi,\ c_a$ see in Appendix \ref{aSRM}.
 Then we can find spectral density of phase quadrature:
 \begin{align}
S_{\phi} &= \frac{\gamma_1\gamma_0 + \gamma_-^2+\Omega^2}{\gamma_+^2+\Omega^2}+\nonumber\\&+\mathcal{H}^2\frac{\gamma_0^2\gamma_1}{|Z|^2}\frac{\gamma_+^4}{\left(\gamma_+^2+\Omega^2\right)^2}\left(\gamma_0+\gamma_1\right) +\nonumber\\&+ 
4\mathcal{H}\frac{\gamma_+^2}{\gamma_+^2+\Omega^2}\frac{\gamma_0\gamma_1}{|Z|^2}\frac{4\kappa_m k_B T}{\hbar}.
 \end{align}
 The first term corresponds to the shot noise, the second term --- to QRPN, and the third to thermal noise.

 \begin{figure}
	\includegraphics[width=0.5\textwidth]{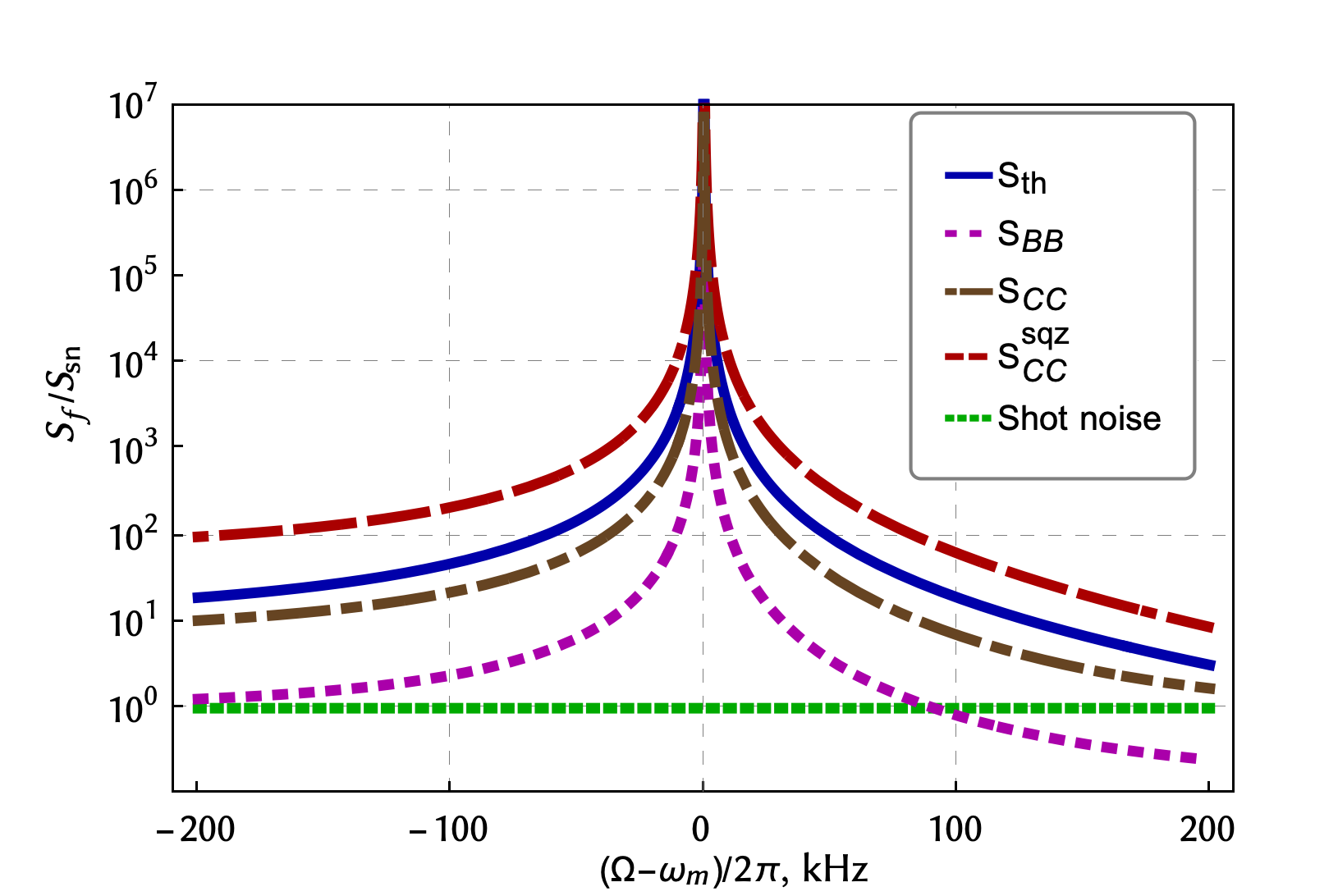}
	\caption{Spectral density normalized to shot noise of optical and thermal fluctuations when measuring the phase quadrature of the output field $C_1$ at $\epsilon=\epsilon_{max}$ \eqref{emax}. 
    Thermal noise $S_{\rm th} $ dominates the sensitivity for  the quality factor of the mechanical oscillator equal to $10^6$. 
    QRPN $S_{CC}$ caused by fluctuations of the field C is just below the thermal noise, and the contribution of laser fluctuations $S_{BB}$ is significantly lower (field $B$ is assumed to be a coherent state).
    By introducing 10\,dB anti-squeezing in the field C, QRPN $S_{CC}^{\rm sqz}$can be increased above the thermal noise level and thus become observable. 
    Alternatively, the quality factor of the mechanical oscillator could be increased by a factor of 10 to reach the same sensitivity.
    Here we assumed the bandwidth of the SRM $\gamma_1/2\pi = 10^6 Hz$ and the MSI $\gamma_0/2\pi = 10^5 Hz$ , the reflection coefficient of the membrane $r_m^2 = 0.98$, see Table\,\ref{Table} for further details.
    }\label{SRMspect}
\end{figure}

Fig.~\ref{SRMspect} shows the spectral density of optical and thermal fluctuations in a table-top setup with SiN membrane as a test mass. In this measurement we measure the phase quadrature of the output field $C_1$ and optimize the imbalance $\epsilon=\epsilon_{max}$ \eqref{emax}. For a typical commercial device the quality factor of the order of $Q=10^6$ results in thermal fluctuations dominating the observed spectrum. In order to observe QRPN, we could either increase Q of the mechanical oscillator by a factor of 10, which is possible for high-quality membranes, or inject 10\,dB of anti-squeezed vacuum into the output port (field $C$). 
For the scheme with PRM, we get the same result, when assuming $\gamma_0\gg\gamma_1$.

\section{Ponderomotive squeezing}
Optomechanical interaction leads to quantum correlation between the mechanical and optical modes\,\cite{braginski1967ponderomotive, khalili2016quantum}. 
They are manifested as reduction in the shot noise level -- ponderomotive squeezing.
 We can use the schemes discussed above to observe this effect. Using a homodyne detector, we measure some quadrature of the output field $C_1$. The natural question is which homodyne angle and which system parameters should we choose for the best observation of the ponderomotive squeezing.

  Fluctuations of the laser light and the thermal noise will greatly interfere with our observation of ponderomotive squeezing. In the case of scheme with SRM (PRM) we will be able to suppress the fluctuations of laser radiation if we take $\gamma_1\gg\gamma_0$ ($\gamma_1\ll\gamma_0$) (see also chapter \ref{BackAction}). An acceptable level of thermal fluctuations is achieved if the Q-factor of the mechanical resonator is $10 ^7$. 

 Let us consider a scheme with SRM. In this case, the quadrature that we will measure has the following form($\theta$ is a homodyne angle):
 \begin{subequations}
 \label{SRMct}
 \begin{align}
c_{1\theta} &= c_{1a}\cos\theta+c_{1\phi}\sin\theta =\\ 
\label{SRMctb}
&= \frac{\gamma_-+i\Omega}{\gamma_+-i\Omega}\left(c_a\cos\theta + c_{\phi}\sin\theta\right) +\\ 
&\quad+ \frac{\sqrt{\gamma_0\gamma_1}}{\gamma_+-i\Omega}\left(b_a\cos\theta + b_{\phi}\sin\theta\right) +\\ 
&\quad+\sqrt{\frac{m}{\hbar}}\frac{\sqrt{\gamma_0\gamma_1}}{\gamma_+-i\Omega}\left(\gamma_- \mathcal{X} \cos\theta - \gamma_+ \mathcal{H} \sin\theta\right)x, 
\end{align}
 \end{subequations}
 where mirror displacement $x$ is equal to
 \begin{subequations}
 \label{SRMct2}
 \begin{align}
 \label{SRMct2a}
x &= -\sqrt{\frac{\hbar}{m}}\frac{\sqrt{\gamma_0\gamma_1}\gamma_+}{\gamma_+-i\Omega}
\frac{\sqrt{\mathcal{X}^2+\mathcal{H}^2}}{Z}\times\\
\label{SRMct2b}
& \qquad \times\left(c_{a}\cos\chi+c_{\phi}\sin\chi\right)-\\ &-\sqrt{\frac{\hbar}{m}}\frac{\gamma_0\gamma_+}{(\gamma_+-i\Omega)Z}\left(\mathcal{H}b_a+i\frac{\Omega}{\gamma_+}\mathcal{X}b_{\phi}\right) + \\
 & \qquad +\frac{F_T}{mZ}, \quad 
\label{chi}
\tan\chi =\frac{\mathcal{X}}{\mathcal{H}}.
 \end{align}
 \end{subequations}

The first terms (\ref{SRMct2a}, \ref{SRMct2b}) describe back action caused by fluctuations of the field $C$. In the case of $ \theta = \chi $ we get a complete correlation of the back action noise and the measurement noise for the field C. We substitute Eq.\eqref{SRMct2} into Eq.\eqref{SRMct} at $\theta = \chi$, keeping only the terms proportional to field C:
\begin{align}
c_{1\theta} &= \left(\frac{\gamma_-+i\Omega}{\gamma_+-i\Omega} + \frac{\gamma_0^2\gamma_1\gamma_+\mathcal{X}\mathcal{H}}{Z(\gamma_+-i\Omega)^2}\right)c_{\theta} +... =\\
\label{SRMcth3}
&= \frac{\gamma_-+i\Omega}{\gamma_+-i\Omega}\frac{\left(\omega_m^2+\frac{\gamma_0^2\gamma_+\mathcal{X}\mathcal{H}}{\gamma_-+i\Omega} - \Omega^2 - i\Omega\kappa_m\right)}{\left(\omega_m^2-\frac{\gamma_0^2\gamma_+\mathcal{X}\mathcal{H}}{\gamma_+-i\Omega}- \Omega^2 - i\Omega\kappa_m\right)}c_{\theta} +...
 \end{align}


Eq.\eqref{SRMcth3} shows that the measuring noise and back action compensate for each other at a certain frequency $\omega_{sq}$ (real part of term in round brackets in nominator is close to zero). Near this frequency, we can observe ponderomotive squeezing of  the quadrature $c_{1\theta}$. In other words, near the frequency $\omega_{sq}$, the power spectral  density of the quadrature $c_{1\theta}$ has a dip with a bandwidth $\Gamma_{sq}$ relative to the level of vacuum fluctuations. The frequency $\omega_{sq}$ and the bandwidth $\Gamma_{sq}$  are equal to
\begin{align}
 \label{wsq}
\omega_{sq} &\simeq \sqrt{\omega_m^2 + \Re\left(\frac{\gamma_+\gamma_0^2\mathcal{X}\mathcal{H}}{\gamma_-+i\omega_m}\right)}= \sqrt{\omega_m^2 + \frac{\gamma_+\gamma_-\gamma_0^2\mathcal{X}\mathcal{H}}{\gamma_-^2+\omega_m^2}},\\
\label{GammaSRM}
\Gamma_{sq} &\simeq \kappa_m + \frac{\Im\left(\frac{\gamma_+\gamma_0^2\mathcal{X}\mathcal{H}}{\gamma_-+i\omega_m}\right)}{\omega_m} = \kappa_m + \frac{\gamma_+\gamma_0^2\mathcal{X}\mathcal{H}}{\gamma_-^2+\omega_m^2}.
 \end{align}

In addition to the ponderomotive squeezing, the power spectral density of  the quadrature $c_{1\theta}$ experiences resonant amplification at a frequency $\omega_M$ with a bandwidth $\Gamma_M$. The frequency $\omega_M$ and the bandwidth $\Gamma_{M}$ are set Eq.\eqref{OmSRM} and Eq.\eqref{OmM}. This resonant amplification can prevent us from observing the ponderomotive squeezing. In order for this not to happen, the following condition must be met:
\begin{align}
\label{eq1}
\frac{\omega_{sq}-\omega_{M}}{\Gamma_M} \simeq \frac{\gamma_1}{2\omega_m} > 1.
 \end{align}

 Also, for the best observation of the ponderomotive squeezing, it is necessary that the bandwidth of the power spectral density dip $\Gamma_{sq}$ is wide enough. Eq.\eqref{GammaSRM} shows that $\Gamma_{sq} \sim \mathcal{X}\mathcal{H}$. In the chapter \ref{BA} we got that the $\mathcal{X}\mathcal{H} \sim \xi\eta$ takes the maximum value at $\epsilon = \epsilon_{opt}$ \eqref{coolingmax}. 
 
 That is, in the case of the scheme with SRM, for the best observation of the ponderomotive squeezing, we must measure the quadrature of the field $C_1$ with a homodyne angle $\theta = \chi$, where $\chi$ is given by Eq\eqref{chi}, the beam splitter coefficient of asymmetry  $\epsilon=\epsilon_{opt}$ \eqref{coolingmax} and $\gamma_1/2 > \omega_m$ \eqref{eq1}.

Let's consider the scheme with PRM. In this case, the quadrature that we will measure has the following form ($\theta$ is a homodyne angle):
 \begin{subequations}
 \label{PRMct}
 \begin{align}
c_{1\theta} &= c_{1a}\cos\theta+c_{1\phi}\sin\theta =\\ 
\label{PRMctb}
&= \frac{-\gamma_-+i\Omega}{\gamma_+-i\Omega}\left(c_a\cos\theta + c_{\phi}\sin\theta\right)+\\ 
&\quad+ \frac{\sqrt{\gamma_0\gamma_1}}{\gamma_+-i\Omega}\left(b_a\cos\theta + b_{\phi}\sin\theta\right) +\\ 
&\quad+\sqrt{\frac{m}{\hbar}}\frac{\sqrt{\gamma_0\gamma_1}}{\gamma_+-i\Omega}\left((\gamma_--i\Omega) \mathcal{X} \cos\theta - \gamma_+ \mathcal{H} \sin\theta\right)x, 
\end{align}
 \end{subequations}
 where mirror displacement $x$ is equal to
 \begin{subequations}
 \label{PRMct2}
 \begin{align}
x &= \sqrt{\frac{\hbar}{m}}\frac{\sqrt{\gamma_0\gamma_1}\gamma_+}{\gamma_+-i\Omega}
\frac{\sqrt{\mathcal{X}^2+\mathcal{H}^2}}{Z} \times\\
\label{PRMct2b}
 &\qquad \times\left(c_{\phi}\cos\beta-c_{a}\sin\beta\right)-\\ &-\sqrt{\frac{\hbar}{m}}\frac{\sqrt{\gamma_0\gamma_1}\gamma_+}{(\gamma_+-i\Omega)Z}\left(\sqrt{\frac{\gamma_1}{\gamma_0}}\mathcal{H}b_a+i\frac{\Omega}{\gamma_+}\mathcal{X}c_{\phi}\right) + \\
 & +\frac{F_T}{mZ}, \quad 
\label{beta}
\tan\beta =\frac{\mathcal{H}}{\mathcal{X}}.
 \end{align}
 \end{subequations}
In the case of $ \theta = \beta +\pi/2 $ we get a complete correlation of the back action noise and the measurement noise for the field C. We substitute Eq.\eqref{PRMct2} into Eq.\eqref{PRMct} at $\theta = \beta +\pi/2 $, keeping only the terms proportional to field C:
\begin{align}
&c_{1\theta} = \left(\frac{-\gamma_-+i\Omega}{\gamma_+-i\Omega} - \frac{\gamma_0\gamma_1\gamma_+\mathcal{X}\mathcal{H}(\gamma_1-i\Omega)}{Z(\gamma_+-i\Omega)^2}\right)c_{\theta} +... =\nonumber\\
\label{PRMcth3a}
&= \frac{-\gamma_-+i\Omega}{\gamma_+-i\Omega}\times\\
\label{PRMcth3b}
&\times\frac{\left(\omega_m^2+\frac{\gamma_0\gamma_1\gamma_+^2\mathcal{X}\mathcal{H}}{(\gamma_--i\Omega)(\gamma_+-i\Omega)} - \Omega^2 - i\Omega\kappa_m\right)}{\left(\omega_m^2-\frac{\gamma_0^2\gamma_+\mathcal{X}\mathcal{H}}{\gamma_+-i\Omega}- \Omega^2 - i\Omega\kappa_m\right)}c_{\theta} +...
 \end{align}

Equations (\ref{PRMcth3a}, \ref{PRMcth3b}) show that for the PRM cavity, there is also a frequency $\omega_{sq}$ near which the measuring noise and the reverse effect begin to compensate for each other. That is, power spectral density of $c_{1\theta}$ has a dip near the frequency $\omega_{sq}$  with a bandwidth $\Gamma_{sq}$. These frequency $\omega_{sq}$ and bandwidth $\Gamma_{sq}$ are equal to
\begin{align}
 \label{wsqprm}
\omega_{sq} &\simeq \sqrt{\omega_m^2 + \Re\left(\frac{\gamma_0\gamma_1\gamma_+^2\mathcal{X}\mathcal{H}}{(\gamma_--i\omega_m)(\gamma_+-i\omega_m)}\right)} =\nonumber\\&= \sqrt{\omega_m^2 + \frac{\gamma_0\gamma_1\gamma_+^2(\gamma_+\gamma_--\omega_m^2)\mathcal{X}\mathcal{H}}{(\gamma_-^2+\omega_m^2)(\gamma_+^2+\omega_m^2)}}, \\
\label{GPRM}
\Gamma_{sq} &\simeq \kappa_m + \frac{1}{\omega_m}\Im\left(\frac{\gamma_0\gamma_1\gamma_+^2\mathcal{X}\mathcal{H}}{(\gamma_--i\omega_m)(\gamma_+-i\omega_m)}\right).
 \end{align}

In order for the resonant amplification not to interfere with the ponderomotive squeezing, the following condition must be met
\begin{align}
\label{eqPRM}
\frac{|\omega_{sq}-\omega_{M}|}{\Gamma_M} \simeq \frac{\gamma_1}{2\omega_m\left(1+\frac{4\omega_m^2}{\gamma_0^2}\right)} > 1.
\end{align}
To obtain this condition, we used Eq.\eqref{OmPRM}, Eq.\eqref{wsqprm} and took into account that $\gamma_1\ll\gamma_0$.

As in the case of SRM $\Gamma_{sq}^{PRM} \sim \mathcal{X}\mathcal{H}$ \eqref{GPRM} reaches its maximum value when $\epsilon = \epsilon_{opt}$ \eqref{coolingmax}.
That is, in the case of the scheme with PRM, for the best observation of the ponderomotive squeezing, we must measure the quadrature of the field $C_1$ with a homodyne angle $\theta = \beta + \pi/2$, where $\beta$ is given by Eq\eqref{beta}, the beam splitter coefficient of asymmetry  $\epsilon=\epsilon_{opt}$ \eqref{coolingmax} and $\gamma_1 > 2\omega_m\left(1+\frac{4\omega_m^2}{\gamma_0^2}\right)$ \eqref{eqPRM}.

\begin{figure}
	\includegraphics[width=0.5\textwidth]{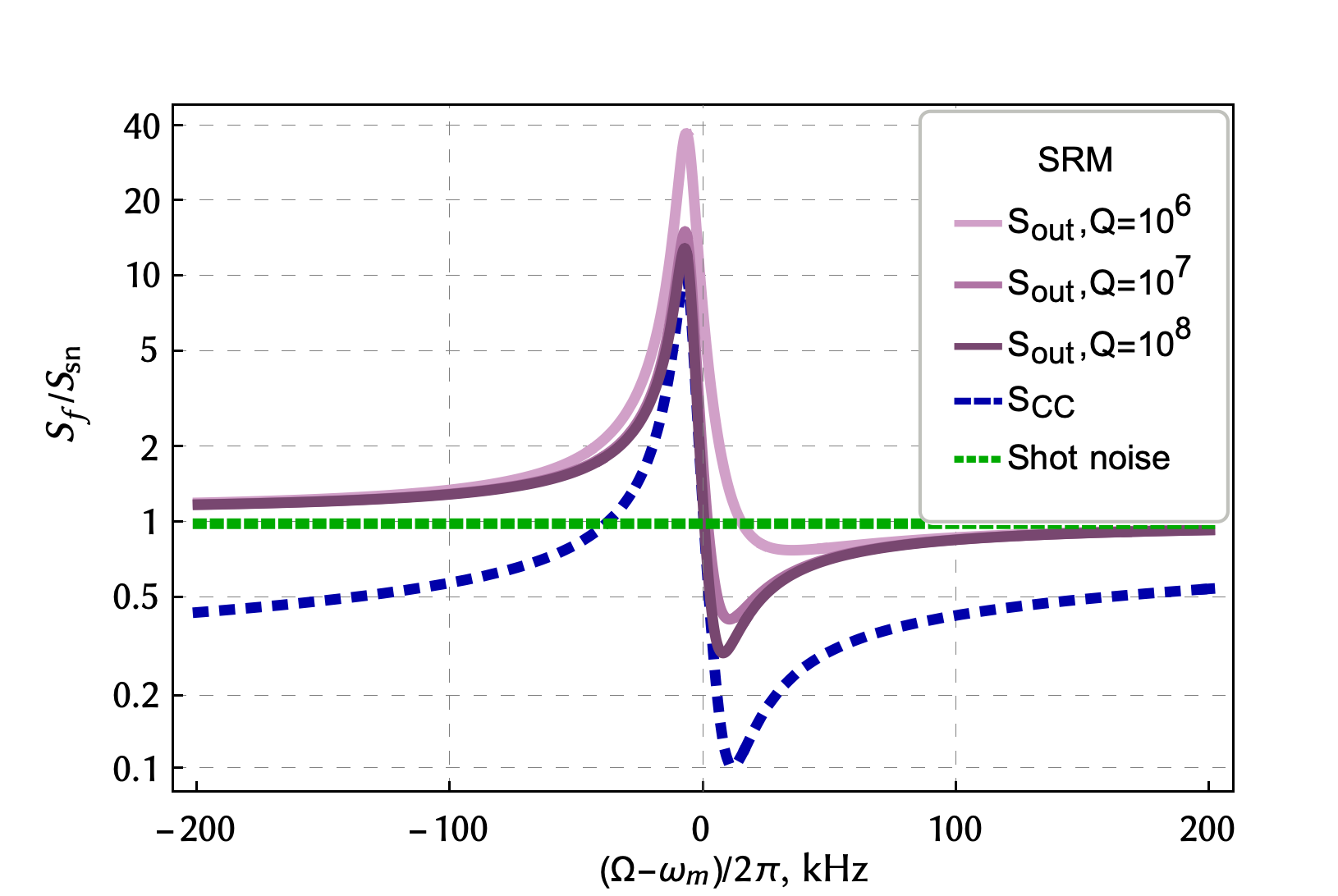}
    \includegraphics[width=0.5\textwidth]{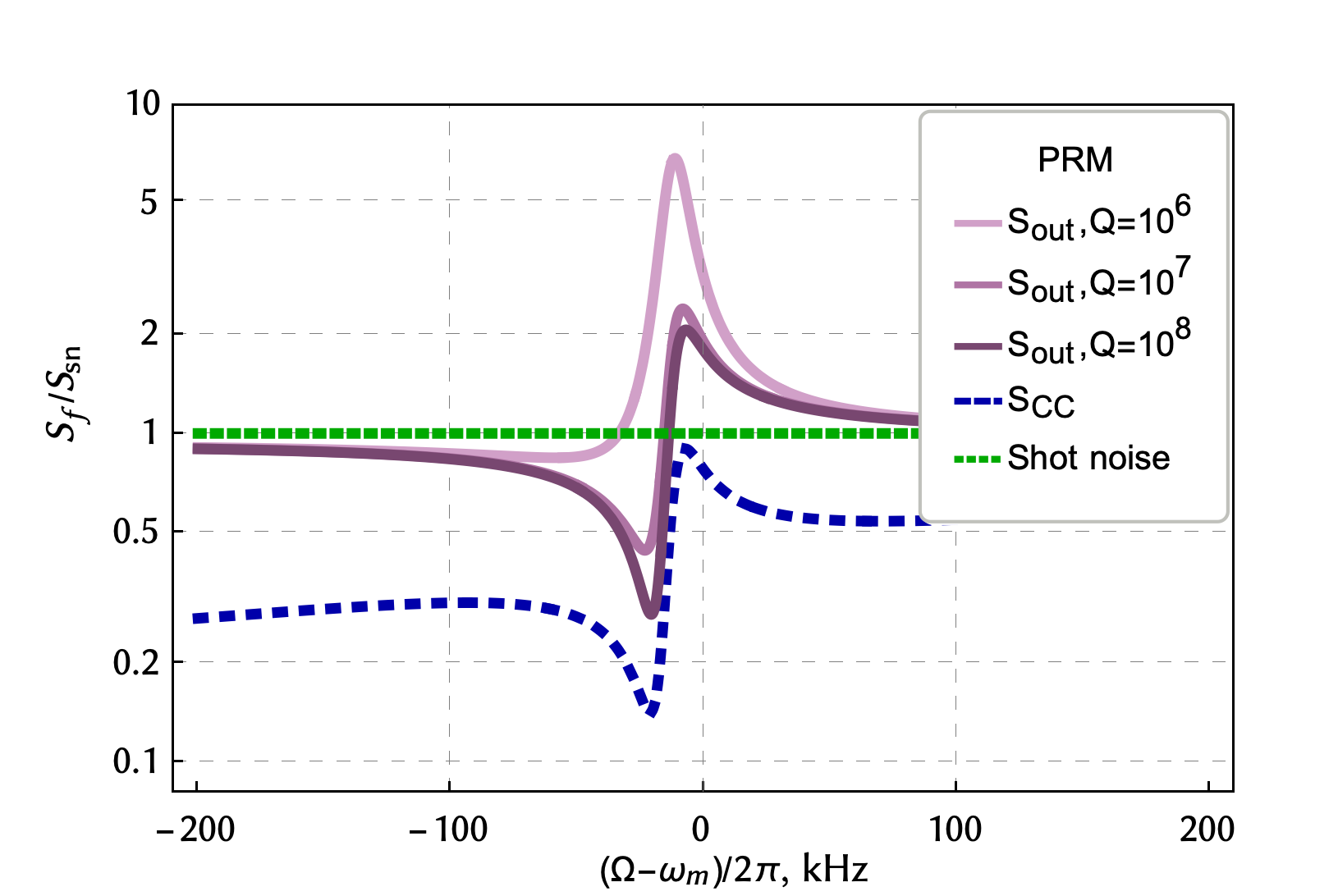}
	\caption{Demonstration of ponderomotive squeezing for different quality of mechanical oscillator. The power spectral densities are normalized to shot noise of a quadrature $c_{1\theta}$ \eqref{SRMct}, \eqref{PRMct} of the output field $C_1$ for SRM (top) and PRM (bottom) schemes. $S_{out}$ is the total noise, including the contribution of thermal noise, as well as the fluctuations of the laser field, and $S_{CC}$ is a contribution of the fluctuation of the field C. Thermal noise prevents observation of strong squeezing, and thus has to be reduced, e.g. by increasing the quality factor of the oscillator. The higher it is, the higher is ponderomotive squeezing, due to reduced contribution of thermal noise. For these plots we used the following parameters:
 (SRM, top) the bandwidth of the SRM $\gamma_1/2\pi = 10^6 Hz$ and the MSI $\gamma_0/2\pi = 3*10^5 Hz$; (PRM, bottom) the bandwidth of the PRM $\gamma_1/2\pi = 3*10^5 Hz$ and the MSI $\gamma_0/2\pi = 10^6 Hz$. The beam splitter asymmetry coefficient $\epsilon=\epsilon_{opt}$ \eqref{coolingmax}, the reflection coefficient of the membrane $r_m^2 = 0.98$ and the values of other parameters are taken from the Table \ref{Table}.}
 \label{SRMsqueezing}
\end{figure}

Fig.~\ref{SRMsqueezing} shows us the power spectral density of a quadrature $c_{1\theta}$ \eqref{SRMct} of the field $C_1$. We see that thermal fluctuations and fluctuations of the laser field quite strongly interfere with the observation of the ponderomotive squeezing. We can improve the situation by using the antisqueezed quadrature $c_{\theta}$ of the field $C$ \eqref{SRMcth3}.

\section{Discussion and Conclusion}
In this paper we presented a novel approach to enhancing the optomechanical coupling in the MSI by unbalancing the central beam splitter.
In this case two fields $A_e$ and $A_n$ (see Fig.~\ref{CavSRM}) inside the interferometer arms have different amplitudes, and they interfere on the beam splitter only partially. 
Because of this, dissipative coupling $\eta \sim \sqrt{1-\epsilon^2}$ decreases (not all light interferes on the beam splitter), dispersive coupling $\xi \sim \epsilon$ increases and a phase-modulated part of the field appears at the MSI output. 

The second mechanism observed in this setup is associated with partial transmission of the movable mirror. 
Components of fields $A_e$ and $A_n$ that have passed through the mirror $M$ appear in the arms of MSI. 
They do not carry information about the displacement of the mirror $M$ in their phases, unlike the reflected parts of the fields. 
Because of this, an interference of an amplitude-modulated signal with an unmodulated signal, which is shifted relative to it by $\pi/2$, occurs at the output MSI. 
This leads to the appearance of dispersive coupling. In this case the increase in dispersive coupling due to the use of an imbalanced beam splitter is more signficant, since the dispersive coupling strength scales as $\xi\sim t_m|T_{\rm MSI}|$, and $\eta \sim r_m$, where $|T_{\rm MSI}|\ll1$.

Dispersive coupling is often used in table-top experiments, such as Fabry-Perot cavities with a movable end mirror or a membrane-in-the-middle.
Using interferometer instead of a single cavity opens several opportunities. 
First, injection of squeezed light is more straightforward, since the signal and laser pump paths are separated\,\cite{kleybolte2020squeezed}.
Second, balance between dispersive and dissipative coupling allows efficient cooling of mechanical motion even on cavity resonance, while in a single cavity some detuning is always required.
In the MSI a small detuning can increase cooling efficiency even further, as it was shown in\,\cite{SawadskyPRL2015}, but the goal of current paper was to study pure effects of the beam-splitter imbalance, leaving detuning as the direction of futre work.
Third, both signal strength and laser power can be tuned independently by introducing signal and/or power recycling mirrors, which opens new level of flexibility.
At the same time, it means that achieving the same order of optomechanical coupling requires both signal and power recyling mirrors to be present.
In this case, however, imbalanced beam-splitter works in a more complex way, and understanding this regime will be the direction of future work.
Fourth, partial interference at the beam-spllitter allows to lower the contribution of classical laser noise to the sensitivity of the setup.
This gives an additional degree of freedom compared to the standard Fabry-Perot cavity.

Compact interferometers with macroscopic test masses serve as test beds for future gravitational-wave detectors, such as Einstein Telescope\,\cite{Punturo2010a} or Cosmic Explorer\,\cite{Reitze2019a}. They allow to test operation of heavy mirrors and their suspensions and develop new technologies for them, e.g. for cryogenic operation\,\cite{utina2022etpathfinder}. MSI offers a compact way to approach such tests, avoiding the need of two separate test masses and allowing for more stable control and alignment of the system, which is especially relevant for cryogenic or quantum-enhanced experiments\,\cite{kleybolte2020squeezed}. The proposed enhancement to optomechanical coupling will allow to further increase the versatility of MSI for research in various aspects of large-scale experiments.

MSI setup provides unusual wealth of optomechanical effects, due to the interplay of dissipative and dispersive optomechanical couplings. 
The new unusual way of operating the interferometer with unbalanced beam-splitter opens even more perspectives for using it to bring macroscopic test masses into quantum domain.
Our work generalises previously studied approaches where the imbalance was absent or very small\,\cite{SawadskyPRL2015, 20PRAKaVy, 22PRAKaVy}, and serves as the next step in understanding the potential benefits of the approach for quantum technology and fundamental search for signatures of quantum gravity, divide between classical and quantum world, and quantum foundations in general.

 \acknowledgments
 A.K. an S.P.V. are grateful for support by Theoretical Physics and Mathematics Advancement Foundation “BASIS” (Grant No. 22-1-1-47-1), by the Interdisciplinary Scientific and Educational School of M.V. Lomonosov Moscow State University ``Fundamental and Applied Space Research''  and by TAPIR GIFT MSU Support from the California Institute of Technology. 
MK was supported by the Deutsche Forschungsgemeinschaft (DFG) under Germany's Excellence Strategy EXC 2121 ``Quantum Universe''-390833306. This document has LIGO number P2300109.

\appendix


\section{SRM cavity}\label{aSRM}

The description of  FP cavity is known (for example, see \cite{Walls2008}). Here we consider SRM cavity shown on Fig.~\ref{CavSRM}, using generalization for cavity with combined dispersive and dissipative coupling and with two ports. The Hamiltonian of such a system can be written as
\begin{subequations}
	\label{Hfull}
\begin{align}
\hat{H} &=\hbar\omega_0(1+\xi\hat{y})\hat{a}_c^{\dagger}\hat{a}_c
    +\frac{\hat{p}^2}{2m}+ \frac{m\omega_m^2 \hat x^2}{2}+\\
   &\qquad  +\hat{H}_{\gamma}+\hat{H}_{T}-F_s\hat{x}.
\end{align}
\end{subequations}
Here $\hat x,\ \hat{p}$ is the coordinate and  momentum of the test mass, $\hat{a}_c$ $\hat{a}_c^{\dagger}$ are annihilation and creation operators describing the intracavity optical field, Hamiltonian $H_T$
\begin{subequations}
\label{HT}
\begin{align}
\hat{H}_{T} &=\int\limits_0^\infty\hbar\omega\, \hat{b}(\omega)^{\dagger}\hat{b}(\omega)\frac{d\omega}{2\pi}
    + \int\limits_0^\infty\hbar\omega\, \hat{c}(\omega)^{\dagger}\hat{c}(\omega)\frac{d\omega}{2\pi}+\\
    &\qquad + \int\limits_0^\infty\hbar\omega\, \hat{g}(\omega)^{\dagger}\hat{g}(\omega)\frac{d\omega}{2\pi},
\end{align}
\end{subequations}
describes two optical thermostats ($\hat b_\omega, \hat c_\omega$ describe input amplitudes in ports $B$ and $C$ correspondingly)  and one mechanical thermostat ($\hat g_\omega$ describe thermal fluctuation force acting on membrane).
Hamiltonian $\hat H_\gamma$ describes optical and mechanical relaxation, for SRM scheme it has form:
\begin{subequations}
\label{Hgamma}
\begin{align}
\hat{H}_{\gamma}^\text{SRM} &=\frac{\hslash}{i} \sqrt{\gamma_0}\left(1+\frac{\eta}{2}\hat{x}\right)
   \left(\hat{b}_{in}\hat{a}_c^{\dagger}-\hat{a}_c\hat{b}_{in}^{\dag}\right)+
   \\
 &\qquad +\frac{\hslash}{i} \sqrt{\gamma_1}     	 
  \left(\hat{c}_{in}\hat{a}_c^{\dagger}-\hat{a}_c\hat{c}_{in}^{\dag}\right)
 +\\
 &\qquad +\frac{\hslash}{i}\sqrt{\kappa_m} \int\limits_0^\infty\left(\hat{g}(\omega)\hat{d}^{\dagger}-\hat{d}\hat{g}(\omega)^{\dag}\right)\,
 \frac{d\omega}{2\pi},\\
  b_{in} &= \int_0^\infty \hat{b}(\omega)e^{-i\omega t} \frac{d\omega}{2\pi},\quad 
  c_{in}=\int_0^\infty \hat{c}(\omega)e^{-i\omega t} \frac{d\omega}{2\pi}, \nonumber
\end{align}	
 \end{subequations}
where $\kappa_m$ is relaxation rate of mechanical oscillator, $d,\ d^\dag$ are its annihilation and creation operators. 
 

From the Hamiltonian \eqref{Hfull} we obtain the set of equations describing the time evolution of the
optomechanical system and 
presenting the annihilation operators of the input and intracavity optical field through slow amplitudes as
\begin{align}
\label{eq2}
\hat{a}_c(t) &\Rightarrow\hat{a}_c(t)e^{-i\omega_0t}, \quad \hat{b}_{in}(t)\Rightarrow\hat{b}_{in}(t)e^{-i\omega_0 t},\\
\hat{c}_{in}(t) &\Rightarrow\hat{c}_{in}(t)e^{-i\omega_0 t}.
\end{align}
we get for slow amplitude in cavity
 \begin{align}
 \label{ac}
  \dot a_c &= -i\omega_0 \xi x\, a_c -\frac{\gamma+\gamma_1}{2}\, a_c 
	 +\sqrt{\gamma}\, b_{in}+\sqrt{\gamma_1}\, c_{in}.
 \end{align}
 For output amplitudes $b_{out},\, c_{out}$ we use well known relations
 \begin{subequations}
 \label{Out1}
 \begin{align}
  b_{out} &= -b_{in} + \sqrt{\gamma}\, a_c,\\
  c_{out} &= -c_{in} +  \sqrt{\gamma_1}\, a_c
 \end{align}
 \end{subequations}
 We apply successive approximations methods and present amplitudes as sum of mean amplitude and small (fluctuation plus signal) assuming that pump light input through port $B$ 
 \begin{subequations}
 \begin{align}
  a_c &= A + a,\quad b_{in} = B+b ,\\
  c_{in}&= c,\quad b_{out} = B_1+b_1,\quad c_{out}=C_1 +c_1
 \end{align}
\end{subequations}
 The fluctuation part of slow amplitude we present using Fourier transform
\begin{subequations}
 \begin{align}
\hat b(t) &= \int_{-\infty}^\infty b(\Omega) \, e^{-i\Omega t}\, \frac{d\Omega}{2\pi}
\end{align}
and similarly for other values denoting the Fourier transform by the same letter without hat. The following commutators and correlators are valid for the Fourier transform of the input fluctuation operators:
\begin{align}
\label{comm11}
\left[ b(\Omega),  b^\dag(\Omega')\right] &= 2\pi\,\delta(\Omega -\Omega'),\\
\label{corr11}
\left\langle b(\Omega)  b^\dag(\Omega')\right\rangle &= 2\pi\, \delta(\Omega -\Omega').
\end{align}
For thermal fluctuation operator acting on mechanical oscillator we have
\begin{align}
  \left[g(\Omega),\, g^\dag(\Omega')\right] &= 2\pi\,\delta(\Omega-\Omega'),\\
  \left\langle g(\Omega)\, g^\dag(\Omega')\right\rangle &=
    2\pi\left(n_T +1\right)\delta(\Omega-\Omega'),\\
  n_T &= \frac{1}{e^{\hslash\Omega/k_B T}-1 } 
 \end{align}
\end{subequations}
Here $n_T$ is a thermal number of quanta.

Below we assume resonance pump. Then for mean amplitudes (zero order of smallness) we have
\begin{subequations}
 \begin{align}
 \label{AB1}
  A  &= \frac{2 \sqrt \gamma_0\, }{\gamma_0 +\gamma_1}\, B,\quad 
  B_1 =\frac{\gamma_0-\gamma_1}{\gamma_o+\gamma_1}\, B,\\
    C_1 & = \frac{2\sqrt{\gamma_0\gamma_1}}{\gamma_1+\gamma_0}\, B
 \end{align}
\end{subequations}
Below we assume that all mean amplitudes  $B,\ A,\ C_1,\ B_1$ are real values.

Using \eqref{ac} for fluctuation amplitude $a$ inside cavity we obtain in frequency domain
\begin{subequations}
 \begin{align}
 \label{aa}
  a &= \frac{ \frac{- i\omega_0 \xi}{\gamma_+} + \frac{\eta \gamma_-}{2\gamma_+}}{\gamma_+ - i\Omega}
	 \cdot\sqrt{\gamma_0}\, x\, B
	 +\frac{\sqrt{\gamma_0} \, b+\sqrt{\gamma_1}\, c}{\gamma_+ - i\Omega},
\end{align}
where we use notations \eqref{gammaPM}.
And we write down equation for position $x$ of test mass
\begin{align}
\label{xx}
  x &Z = -\frac{\xi\, \hslash\omega_0}{m} \, A\left(a^\dag +a\right) -\\
     &\quad  - \frac{i\hslash \eta\sqrt{\gamma_0}}{2m}\left(B a^\dag  + b A^* -B^* a - A b^\dag\right) + \frac{F_s}{m}=\nonumber\\
    &= -\frac{\sqrt 2 \xi\, \hslash\omega_0 A}{m} \, a_a -\\
    &\quad  - \frac{\hslash \eta \sqrt{2\gamma_0}}{2m}\left(B a_\phi  -  A b_\phi\right)  +\frac{F_s}{m},
\end{align}
where we introduce notation for impedance $Z$ and quadrature amplitudes
\begin{align}
    Z & = \left(\omega_m^2-\Omega^2 -i\Omega \kappa_m\right),\\
    a_a &=\frac{a(\Omega)+a^\dag(-\Omega)}{\sqrt 2},\quad 
     a_\phi=\frac{a(\Omega)-a^\dag(-\Omega)}{i\sqrt 2},
\end{align}
and similarly for $b_a,\ b_\phi$.
\end{subequations}

Force acting on probe mass $m$ can be divided into four part --- optical rigidity force $F_x$, back action fluctuations force $F_{ba}$, thermal fluctuations $F_T$ and signal force $F_s$.

In order to find optical rigidity we take terms in \eqref{aa}, depending on position $x$, and substitute into \eqref{xx}. As result we obtain optical rigidity force
\begin{subequations}
 \begin{align}
 \frac{F_x}{m}	& = -\omega_{os}^2 x, \quad 
	\omega_{os}^2=- \frac{\hslash \omega_0 \gamma_0^2 \eta \xi B^2}{m\gamma_+^2(\gamma_+ - i\Omega)}.
\end{align}
\end{subequations}
It is coincides with \eqref{OmSRM} if we account 
\begin{align}
\label{Win}
W_{in}= \hslash \omega_0 B^2.
\end{align}
Note that $\omega_{os}^2$ is a {\em complex} value: real part corresponds to rigidity, whereas imaginary part --- introduced relaxation.

Fluctuation part of back action fluctuation force (light pressure):
\begin{subequations}
 \begin{align}
  \label{Ffl}
  \frac{F_{fl}}{m} &= -\frac{\sqrt 2 \xi\, \hslash\omega_0 \sqrt{\gamma_0} B}{m \gamma_+} \cdot
     \frac{\sqrt{\gamma_0} \, b_a+\sqrt{\gamma_1}\, c_a}{\gamma_+ - i\Omega}-\\
	&\quad - \frac{\hslash \eta \sqrt{2\gamma_0}B}{2m}
	 \left( 
	 \frac{\sqrt{\gamma_1}\, c_\phi}{\gamma_+ - i\Omega}  
	 +  \frac{\sqrt{\gamma_0}\, i\Omega}{\gamma_+(\gamma_+-i\Omega)}\, b_\phi\right).
	 \nonumber
 \end{align}
\end{subequations}
Finally we obtain equation for displacement $x$ 
\begin{subequations}
 \begin{align}
 \label{x2}
  x &= \frac{-\hslash \sqrt{\gamma_0}B}{m Z_m (\gamma_+ -i\Omega)}\left(
  \frac{\sqrt 2 \xi \omega_0}{\gamma_+}\big[\sqrt{\gamma_0} \, b_a+\sqrt{\gamma_1}\, c_a\big]
  +\right.\\
   & \quad \left.+ \frac{\eta}{\sqrt 2}\left[\sqrt{\gamma_1}\, c_\phi 
	 + \frac{\sqrt{\gamma_0}\, i\Omega\, b_\phi}{\gamma_+}\right]\right)
    +\frac{F_s +F_T}{m Z_m},\nonumber
 \end{align}
 where impedance $Z_m$ is modified due to optical rigidity
 \begin{align}
 \label{Zm}
 Z_m &\equiv \left(\omega_M^2-\Omega^2 -i\Omega \kappa_M\right),\\
  f_s& =\frac{F_s}{\sqrt{2\hslash m\omega_M}},
 \quad f_T =\frac{F_T}{\sqrt{2\hslash m\omega_M}},
 \end{align}
 \end{subequations}
 where $\omega_M$ and $\kappa_M$ are defined by \eqref{OmM}.
 
 It is convenient to rewrite \eqref{x2} in form
 \begin{subequations}
  \label{x3}
  \begin{align}
   x =& \frac{\sqrt{2\hslash m\omega_M}}{mZ_m} \times\\
   &\ \left\{-\sqrt\frac{\gamma_0^2}{2\omega_M}\frac{\gamma_+}{ (\gamma_+ -i\Omega)}\left(
  {\mathcal{H}}\left[ \, b_a+\sqrt\frac{\gamma_1}{\gamma_0}\, c_a\right]\right.\right.\nonumber\\
   &\quad \left.\left.+ {\mathcal{X}}\left[\sqrt\frac{\gamma_1}{\gamma_0}\, c_\phi 
	 + \frac{ i\Omega\, b_\phi}{\gamma_+}\right]\right)   +f_s +f_T \right\}.  
  \end{align}
Spectral density $S_{LP}^{SRM}$ in \eqref{SSRM} corresponds to spectral density of term in figure brackets in \eqref{x3}. Here coefficients $\mathcal{X},\ \mathcal{H}$ are defined in \eqref{PQ} with account of \eqref{Win}.
 \end{subequations}

 \section{PRM cavity}\label{aPRM}
 
 Formulas for Hamiltonian (\ref{Hfull}, \ref{HT}) are  valid for PRM cavity, the formula \eqref{Hgamma} has to be rewritten in form
 \begin{subequations}
\label{HgammaP}
\begin{align}
\hat{H}_{\gamma}^\text{PRM} &=\frac{\hslash}{i} \sqrt{\gamma_0}\left(1+\frac{\eta}{2}\hat{x}\right)
   \left(\hat{c}_{in}\hat{a}_c^{\dagger}-\hat{a}_c\hat{c}_{in}^{\dag}\right)+
   \\
 &\qquad +\frac{\hslash}{i} \sqrt{\gamma_1}     	 
  \left(\hat{b}_{in}\hat{a}_c^{\dagger}-\hat{a}_c\hat{b}_{in}^{\dag}\right)
 +\\
 &\qquad +\frac{\hslash}{i}\sqrt{\kappa_m} \int\limits_0^\infty\left(\hat{g}(\omega)\hat{d}^{\dagger}-\hat{d}\hat{g}(\omega)^{\dag}\right),
 \frac{d\omega}{2\pi}.
\end{align}	
\end{subequations}
 The equation \eqref{ac} is also should be rewritten
\begin{align}
 \label{acP}
  \dot a_c &= -i\omega_0 \xi x\, a_c -\frac{\gamma+\gamma_1}{2}\, a_c 
	 +\sqrt{\gamma_1}\, b_{in}+\sqrt{\gamma}\, c_{in}.
 \end{align}
 As well as output amplitudes $b_{out},\, c_{out}$ \eqref{Out1}:
 \begin{subequations}
 \label{Out1P}
 \begin{align}
  b_{out} &= -b_{in} + \sqrt{\gamma_1}\, a_c,\\
  c_{out} &= -c_{in} +  \sqrt{\gamma}\, a_c
 \end{align}
 \end{subequations}

 Mean amplitudes of fields:
 \begin{subequations}
 \begin{align}
 \label{AB2}
  A  &= \frac{2 \sqrt \gamma_1\, }{\gamma_0 +\gamma_1}\, B,\quad 
  B_1 =\frac{\gamma_1-\gamma_0}{\gamma_o+\gamma_1}\, B,\\
    C_1 & = \frac{2\sqrt{\gamma_0\gamma_1}}{\gamma_1+\gamma_0}\, B
 \end{align}
\end{subequations}
 
 Optical rigidity for PRM cavity has also to be corrected \eqref{OmPRM}.

Fluctuation part of back action light pressure force \eqref{Ffl} is also should be coorrected:
\begin{subequations}
 \begin{align}
  \label{FflPRM}
  \frac{F_{fl}}{m} &= -\frac{\sqrt 2 \xi\, \hslash\omega_0 \sqrt{\gamma_1} B}{m \gamma_+} \cdot
     \frac{\sqrt{\gamma_1} \, b_a+\sqrt{\gamma_0}\, c_a}{\gamma_+ - i\Omega}-\\
	&\quad - \frac{\hslash \eta \sqrt{2\gamma_1}B}{2m}
	 \left( 
	 \frac{\sqrt{\gamma_0}\, c_\phi}{\gamma_+ - i\Omega}  
	\right).
	 \nonumber
 \end{align}
\end{subequations}
And also  equation \eqref{x2} for displacement $x$ 
\begin{subequations}
 \begin{align}
 \label{x2PRM}
  x &= \frac{-\hslash \sqrt{\gamma_1}B}{m Z_m (\gamma_+ -i\Omega)}\left(
  \frac{\sqrt 2 \xi \omega_0}{\gamma_+}
    \big[\sqrt{\gamma_1} \, b_a+\sqrt{\gamma_0}\, c_a\big]   +\right.\\
   & \quad \left.+ \frac{\eta}{\sqrt 2}\left[\sqrt{\gamma_0}\, c_\phi\right]\right)
    +\frac{F_s +F_T}{m Z_m},\nonumber
 \end{align}
\end{subequations}
And also we write analog of \eqref{x2} for PRM
\begin{subequations}
 \label{x3PRM}
 \begin{align}
  x =& \frac{\sqrt{2\hslash m\omega_M}}{mZ_M}\\
  & \times\left\{\sqrt\frac{\gamma_1}{2\omega_M}
   \left(  -  \mathcal{H} \cdot
  \frac{\gamma_+\left(\sqrt{\gamma_1} \, b_a+\sqrt{\gamma_0}\, c_a\right)}{\gamma_+ - i\Omega}\right.\right.\\
  & \quad \left.\left.   +  \mathcal{X} \cdot \sqrt{\gamma_0}\cdot c_\phi \right)
           +f_s +f_T \frac{}{}\right\}.
 \end{align}
Spectral density $S_{LP}^{PRM}$ in \eqref{SPRM} corresponds to spectral density of term in figure brackets in \eqref{x3PRM}. Coefficients $\mathcal{X},\ \mathcal{H}$ are defined in \eqref{PQ}.
\end{subequations}

\bibliographystyle{ieeetr}
\bibliography{OptoMech.bib}

\end{document}